\documentclass[aps,pre,twocolumn,superscriptaddress,nolongbibliography]{revtex4-2}

\DeclareRobustCommand*{\citen}[1]{%
  \begingroup
    \onlinecite{#1}%
  \endgroup
}

\usepackage{graphicx}
\usepackage{physics}
\usepackage{amssymb}   
\usepackage{float}
\usepackage[caption=false]{subfig}
\usepackage{comment}
\usepackage{hyperref}
\usepackage{xcolor}

   \let\k=\kappa

\renewcommand{\vec}[1]{\mathbf{#1}}

\def\to{\rightarrow}

\newcommand{\new}{\nonumber\\}

\newcommand{\fdiff}[2]{\fdv{#1}{#2}}
\newcommand{\diff}[2]{\dv{#1}{#2}}

\newcommand{\ave}[1]{\left\langle #1 \right\rangle}
\renewcommand{\abs}[1]{\left| #1 \right|}
\newcommand{\bx}{\vec{x}}
\newcommand{\by}{\vec{R}}
\newcommand{\bX}{\vec{X}}
\newcommand{\ox}{\underline{\bx}}
\newcommand{\oy}{\underline{\by}}

\usepackage{stackengine}
\stackMath
\newcommand\subline[2]{\stackon[-1.5pt]{\hspace{-10pt}#1}{\hspace{4pt}\rule[2pt]{\widthof{$#1$}}{.4pt}_{#2}}}

\begin{document}
\title{Mean-field caging in a random Lorentz gas}

\author{Giulio Biroli}
\affiliation{Laboratoire de Physique de l'Ecole Normale Sup\'erieure, ENS, Universit\'e PSL, CNRS, Sorbonne Universit\'e, Universit\'e de Paris, F-75005 Paris, France
}
\author{Patrick Charbonneau}
\affiliation{Department of Chemistry, Duke University, Durham, North Carolina 27708, USA}
\affiliation{Department of Physics, Duke University, Durham, North Carolina 27708, USA}
\author{Yi Hu}
\email{yi.hu@duke.edu}
\affiliation{Department of Chemistry, Duke University, Durham, North Carolina 27708, USA}
\author{Harukuni Ikeda}
\affiliation{Graduate School of Arts and Sciences, The University of Tokyo 153-8902, Japan
}
\author{Grzegorz Szamel}
\affiliation{Department of Chemistry, Colorado State University, Fort Collins, CO 80523, USA}
\author{Francesco Zamponi}
\affiliation{Laboratoire de Physique de l'Ecole Normale Sup\'erieure, ENS, Universit\'e PSL, CNRS, Sorbonne Universit\'e, Universit\'e de Paris, F-75005 Paris, France
}

\begin{abstract}
The random Lorentz gas (RLG) is a minimal model of both percolation and glassiness, which leads to a paradox in the infinite-dimensional, $d\rightarrow\infty$ limit: the localization transition is then expected to be \emph{continuous} for the former and \emph{discontinuous} for the latter. As a putative resolution, we have recently suggested that as $d$ increases the behavior of the RLG converges to the glassy description, and that percolation physics is recovered thanks to finite-$d$ perturbative and non-perturbative (instantonic) corrections [Biroli \emph{et al}.~arXiv:2003.11179]. Here, we expand on the $d\rightarrow\infty$ physics by considering a simpler static solution as well as the dynamical solution of the RLG. Comparing the $1/d$ correction of this solution with numerical results reveals that even perturbative corrections fall out of reach of existing theoretical descriptions. Comparing the dynamical solution with the mode-coupling theory (MCT) results further reveals that although key quantitative features of MCT are far off the mark, it does properly capture the discontinuous nature of the $d\rightarrow\infty$ RLG. These insights help chart a path toward a complete description of finite-dimensional glasses.
\end{abstract}

\maketitle

\section{Introduction}

Formulating a first-principle description of glasses remains a major challenge of condensed matter and statistical physics. A proposal for a constructive approach is to solve for the mean-field theory of glasses in the limit of infinite spatial dimensions, $d \rightarrow \infty$, and to then introduce systematic finite-$d$ corrections~\cite{kirkpatrick1987connections,kirkpatrick1987stable,charbonneau2017glass}. For standard phase transitions this approach captures the salient behavior of physical systems in $d=2$ and 3, and the relatively smooth dimensional evolution of the glass phenomenology suggests that the same should apply for these richer systems as well~\cite{charbonneau2017glass}. Technically rooted in the celebrated spin glass solution~\cite{mezard1987spin} and in the foundational works of Ted Kirkpatrick, Dave Thirumalai and Peter Wolynes in the mid-1980s~\cite{kirkpatrick1987connections,kirkpatrick1987stable,kirkpatrick1989random,kirkpatrick1989scaling}, a mean-field $d \to \infty$ theory for structural glasses has been formulated in recent years~\cite{parisi2010mean,francesco2020theory}. These works, which follow the idea of using static replica methods to find solutions of random Hamiltonians~\cite{kirkpatrick1989random,kirkpatrick2012random} (see also Refs~\citen{monasson1995structural,mezard1996tentative}), have demonstrated -- in the $d \to\infty$ limit -- that the random first-order transition description of glasses is valid~\cite{kirkpatrick1989scaling}, and that hard spheres are natural archetypes of all simple liquids~\cite{parisi2010mean,charbonneau2017glass,francesco2020theory}. 

The finite-$d$ physics of hard sphere liquids is, however, far from simple. Different non-mean-field effects such as hopping~\cite{CiamarraSM2016}, facilitation~\cite{berthier2011theoretical,BiroliGarrahan2013} and nucleation~\cite{dzero2005activated} are then observed. Their competing effects make the analysis of finite-$d$ corrections rather difficult.
Modified models have thus been introduced in an attempt to isolate some of these effects. For example, systems of hard spheres with randomly shifted pair distances, as proposed by Mari, Kurchan and Krzakala (MKK)~\cite{mari2009jamming,mari2011dynamical}, can systematically eliminate multi-body interactions, while still behaving similarly as hard sphere glasses in the limit $d\rightarrow\infty$~\cite{mezard2011solution,charbonneau2014hopping}. In that vein, we have recently found that an even simpler, single-particle model, the random Lorentz gas (RLG), can be solved in the $d\to\infty$ limit and that its solution shares all the features of the equilibrium hard sphere equivalent~\cite{biroli2020unifying}. Because, by construction, most finite-$d$ corrections are then absent -- including nucleation, structural correlations and facilitation -- those that do persist are especially noticeable.  In particular, although the RLG has long been known as a model of continuum percolation~\cite{kerstein1983equivalence,elam1984critical,hofling2006localization,hofling2008critical,bauer2010localization,spanner2016splitting,petersen2019anomalous,charbonneau2020percolation}, that description and the $d\to\infty$ mean-field one fundamentally disagree about the nature of the localization transition. It should be continuous according to the former, and discontinuous according to the latter. To resolve this paradox, we have proposed that instantonic (non-perturbative) hopping events intervene in finite $d$~\cite{biroli2020unifying}. Because numerical simulations of the RLG can be pushed as high as $d=20$, we also gathered substantial quantitative support for this proposal. 
This analysis, however, leaves many questions open. For instance, it remains unclear what the origin of the sizable perturbative, $1/d$, corrections identified in Ref.~\citen{biroli2020unifying} might be.   

We also note that the mode-coupling theory (MCT) of the RLG predicts a continuous localization transition in finite $d$ whereas the intimately related  MCT of glasses~\cite{gotze1981dynamical,leutheusser1984dynamical,jin2015dimensional} predicts a discontinuous dynamical glass transition. Since both mode-coupling theories are of mean-field flavor and do not include any contribution from hopping processes, one wonders whether the prediction of a continuous localization transition is a happy artefact or whether some mean-field-like effects influence the character of the transition. 

Following Ref.~\citen{charbonneau2020percolation}, in which some of us investigated the percolation caging behavior of the RLG, we here investigate the mean-field caging behavior of the RLG as $d$ increases. The plan for the rest of this work is as follows. We first compare static mean-field solutions obtained by virial expansion by taking the high-asymmetry limit of binary hard spheres mixtures and the infinite-radius limit of the non-convex perceptron.  We then derive the dynamical solution of the $d\to\infty$ RLG. Mean-field predictions are finally compared with numerical cavity reconstruction results, before briefly concluding.

\section{Static derivations}
\label{sec:static}

In Ref.~\citen{biroli2020unifying}, we confirmed that the mean-field RLG belongs to the same universality class as hard sphere glasses in the $d \rightarrow \infty$ limit by an ansatz-free cavity reconstruction calculation. Here, we extend this analogy by considering various mean-field approaches based on different assumptions and limits. We first consider the virial solution of the RLG using a Gaussian cage ansatz, and then investigate the RLG as a limit case of a binary  hard spheres mixture~\cite{coluzzi1999thermodynamics,biazzo2009theory,ikeda2016note}. We additionally consider the possibility of describing the RLG as a special limit of the non-convex perceptron, a model that has been particularly informative about the related physics of jamming~\cite{franz2016simplest,Franz2015,Franz2017,Franz2019}. 

For the sake of clarifying the notation, recall that the RLG consists of an infinitesimally small tracer navigating within the space left by $N$ Poisson-distributed hard spherical obstacles in a $d$-dimensional box of volume~$V$. The unitless obstacle density can thus be given by
\begin{equation} \label{eq:densitydef}
\Phi = \rho V_d \sigma^d
\end{equation}
where $\rho=N/V$ is the number density of the obstacles of radius $\sigma$, and $V_d$ is the $d$-dimensional volume of a ball of unit radius, $V_d = \pi^{d/2}/\Gamma(1+d/2)$. It is worth noting that the RLG physics is invariant to rescaling the obstacle and tracer radii, as long as their sum is fixed to $\sigma$. The RLG can thus also be viewed as a limit case of a binary mixture of hard spheres, with obstacles being infinitely smaller and heavier than the infinitely-dilute tracer. In the $d\to\infty$ limit, the RLG presents a glass-like dynamical caging transition at $\hat\varphi_\mathrm{d}$, whose order parameter is the cage size $\Delta$, defined by the long-time limit of the tracer mean squared displacement (MSD). For convenience, we also define the rescaled density $\hat\varphi = \Phi/d$ and cage size $\hat\Delta = d \Delta$, which reach a finite limit at the caging transition when $d\to\infty$.

The RLG partition function is simply the total free available volume. For a specific realization of quenched disorder, i.e., obstacle positions, it can be written as
\begin{align}\label{eq:Z}
Z = \int \dd \bx \prod_{i=1}^N \theta(\abs{\bx-\by_i}-\sigma) \ , 
\end{align}
where the Heaviside step function $\theta$ enforces volume exclusion by the obstacles, and the $d$-dimensional vectors $\bx$ and $\by_i$ denote the positions of the tracer and of the $i$-th obstacle, respectively. Assuming that the free-energy is self-averaging and using the replica method, we then get 
\begin{align}
    -\beta F &= \lim_{n\to 0}\frac{\log\overline{Z^n}}{n} \ ,\\
    \label{eq:replicaZ}
\overline{Z^n} &= \int \dd \ox \prod_{i=1}^N
     \overline{\prod_{a=1}^n \theta(\abs{\bx^a-\by_i}-\sigma)} \ ,
     \nonumber
\end{align}
where $\ox=\{\bx^1,\cdots, \bx^n \}$ denotes a set of $n$ identical replicas of the original tracer, and $\overline{\bullet}$ denotes averaging over obstacle positions $\by_i$.

\subsection{Virial expansion}
\label{sec:virial}
We first consider a solution of the RLG using a liquid-state approach. As is canonical in that context, we introduce an external field 
$\psi(\ox)$ as~\cite{hansen1990theory}
\begin{align}
\overline{Z^n} \to Z[\psi(\ox)] &\equiv \int \dd \ox e^{\psi(\ox)} \prod_{i=1}^N 
\overline{\prod_{a=1}^n \theta(\abs{\bx^a-\by_i}-\sigma)} \new
&= \int \dd \ox e^{\psi(\ox)}G(\ox),
\end{align}
with auxiliary function
\begin{align}
 G(\ox) &\equiv \prod_{i=1}^N \overline{\prod_{a=1}^n \theta(\abs{\bx^a-\by_i}-\sigma)}.
\end{align}
The field $\psi(\ox)$ is conjugate to the density distribution of the replicas,
\begin{align}
 \rho(\ox) &= \ave{\prod_{a=1}^n \delta(\bx-\bx^a)} = \fdiff{\log Z[\psi(\ox)]}{\psi(\ox)} \new
 &= \frac{1}{Z[\psi]}e^{\psi(\ox)}G(\ox)\label{175901_17Oct17},
\end{align}
and hence
\begin{align}
 \log\rho(\ox) &= \psi(\ox) + \log G(\ox) -\log Z[\psi].
\end{align}
From this relation, we obtain a density functional~\cite{singh1985hard,kirkpatrick1989random} for $\rho(\ox)$
\begin{align}
 \Omega[\rho] &\equiv \log Z[\psi]-\int \dd \ox \rho(\ox)\psi(\ox)\new
 &= -\int \dd \ox \rho(\ox)\log\rho(\ox) + \int \dd \ox \rho(\ox)\log G(\ox)\new
 &\approx -\int \dd \ox \rho(\ox)\log\rho(\ox) + \rho \int \dd \ox \rho(\ox) \int \dd\by f(\ox,\by),\label{182206_17Oct17}
\end{align}
with Mayer function
\begin{align}
 f(\ox,\by) &= \prod_{a=1}^n \theta(\abs{\bx^a-\by}-\sigma) -1.
\end{align}
Note that this expression is equivalent to that for binary mixtures~\cite{ikeda2017decoupling}, thus strengthening the liquid-state framework analogy. 

The final step entails optimizing the free-energy with respect to the density distribution, $\rho(\ox)$. In general this operation can be quite challenging, but in the $d\to\infty$ limit the calculation greatly simplifies~\cite{francesco2020theory}. The central limit theorem indeed guarantees that only the second moment of $\rho(\ox)$ matters for the free energy computation. The density $\rho(\ox)$ can thus be taken as a Gaussian distribution, which for a one-replica symmetry breaking (1RSB) ansatz can be written as
\begin{align}
 \rho(\ox) &=
 \prod_{k=1}^{n/m}\frac{1}{V}\int \dd X_k \prod_{b\in B_k}\gamma_{A}(\bx^b-X_k),\label{170446_17Oct17}
\end{align}
where $B_k=\{mk+1,mk+2,\cdots, m(k+1)\}$ is the $k$th replica block, and $\gamma_A(\vec{r})$ denotes a $d$-dimensional Gaussian with zero mean and variance $A$.
Within this ansatz, the free-energy expression becomes
\begin{equation} \label{eq:virialF}
\begin{aligned} 
 -\beta F = &\lim_{n\to 0}\frac{\Omega}{n}  = \frac{1}{m} \Big[\log V \\
 &+ \frac{d}{2}(m-1)\log(2\pi A)  + \frac{d}{2}\log m + \frac{d}{2}(m-1)\Big] \\
 &+ \frac{1}{m}\rho \int \dd \vec{r} [q_{A/2}(\vec{r})^m-1],
\end{aligned}
\end{equation}
where 
\begin{align}
    q_A(\vec{r}) = \int \dd \vec{r}' \gamma_A(\vec{r}-\vec{r}')\theta(\abs{\vec{r}-\vec{r}'}).
\end{align}
For the RLG, the radial distribution function for obstacles is simply a step function, and in that sense is equivalent to the Mari-Kurchan model. Hence, $q_A(\vec{r})$ can be directly taken from Eq.~[S19] of Ref.~\citen{charbonneau2014hopping} as
\begin{equation}
\begin{aligned} \label{eq:qagaussian}
q_A(r) &= \int_\sigma^\infty \dd u \left( \frac{u}{r} \right)^{\frac{d-1}{2} } e^{\frac{(r-u)^2}{4A}}  \frac{\sqrt{ru}}{2A} I_{\frac{d-2}{2} } \left(\frac{ru}{2A}\right),
\end{aligned}
\end{equation}
where $I_n(x)$ is the modified Bessel function, and the vector integration $q_A(\vec{r})$ can be transformed into a scalar integration by using rotational invariance. The saddle point condition for Eq.~\eqref{eq:virialF} finally gives
\begin{align}\label{eq:saddlepointm}
 \frac{1}{A} &= \frac{2\rho}{d(m-1)}\pdv{}{A}\int \dd \vec{r} q_{\frac A 2}(\vec{r})^m.
\end{align}

The dynamical glass transition corresponds to taking the limit $m\to 1$, in which case Eq.~\eqref{eq:saddlepointm} reduces to 
\begin{align} \label{eq:cagesizegaussian}
 \frac{1}{A} &= -\frac{2\rho}{d}\pdv{}{A}\int \dd \vec{r} q_{\frac A 2}(\vec{r})\log q_{\frac A 2}(\vec{r}).
\end{align}
The dynamical glass transition point is then
\begin{align}
 \frac{1}{\rho_\mathrm d} &= -\max_{A}\frac{2A}{d}\pdv{}{A}\int \dd \vec{r} q_{\frac A 2}(\vec{r})\log q_{\frac A 2}(\vec{r}).
 \label{154732_15Feb18}
\end{align}
Identifying $\hat\varphi$ and substituting $\hat\Delta = 2 d^2 A$ in Eq.~\eqref{eq:cagesizegaussian} results in the exact same expression as for the ansatz-free cavity derivation of Ref.~\citen{biroli2020unifying}, thus demonstrating the equivalence of the two approaches in the limit $d\rightarrow\infty$. 

Unlike the cavity derivation, however, this approach also naturally leads to a model for finite-$d$ corrections~\cite{parisi2010mean,mangeat2016quantitative}. Assuming that Gaussian caging holds for all dimensions indeed allows for  Eq.~\eqref{eq:qagaussian} to be solved in any given $d$. Although the ensuing correction to $\varphi_\mathrm{d}$ is marginal~\cite{mangeat2016quantitative}, that for $\hat\Delta$ is more substantial. Results from this model are compared with finite-$d$ numerics in later section.

\subsection{Binary mixture derivation}
\label{sec:binary}
In this subsection we draw a parallel between the RLG and a limit case of a binary hard sphere mixture. Recall from  Ref.~\citen{ikeda2017decoupling} that the free energy of a $n$-replicated binary mixture of large and small particles, $\mu=\{{\rm Large, Small}\}$, is given in the $d\rightarrow\infty$ limit as
\begin{equation} \begin{aligned}
-\beta F &= \sum_\mu 
\int \dd \ox \rho_\mu(\ox)(1-\log\rho_\mu(\ox)) \\
&\quad + \frac{1}{2}\sum_{\mu\nu}\int \dd \ox \dd \oy \rho_\mu(\ox)\rho_\nu(\oy)f_{\mu\nu}(\ox,\oy),
\label{binary}
\end{aligned} \end{equation}
where the density distribution of the replicas is
\begin{subequations}
\begin{align}
\rho_{\rm Large}(\ox) = \sum_{i\in {\rm Large}}\ave{\prod_{a=1}^n \delta(\bx^a-\bx_i^a)},\\
\rho_{\rm Small}(\ox) = \sum_{i\in {\rm Small}}\ave{\prod_{a=1}^n \delta(\bx^a-\bx_i^a)}.
\end{align}
\end{subequations}
In order to model the RLG, we consider a limit that satisfies the following three conditions:
\begin{enumerate}
\item neglect large-large and small-small interactions;
\item freeze large particle positions;
\item include $N-1\sim N$ large particles and one small particle.
\end{enumerate}
Under these constraints, the Mayer functions are 
\begin{subequations}
\begin{align}
&   f_{\rm Large,Large} = f_{\rm Small,Small} =0,\\
&   f_{\rm Large,Small} = f,\\
&    \rho_{\rm Large}(\ox) \to \rho \int \dd\bX\prod_{a=1}^n\delta(\bX-\bx^a),\\
&    \rho_{\rm Small}(\ox) \to \rho(\ox),
\end{align}
\end{subequations}
which upon substitution into Eq.~\eqref{binary} immediately gives
\begin{equation} \begin{aligned}
-\beta F =& \rho \int \dd \ox \rho(\ox)\int \dd \by f(\ox,\by) \\
&\quad -\int \dd \ox \rho(\ox)\log\rho(\ox)+ {\rm cnst}.
\end{aligned} \end{equation}
The resulting expression is the same as Eq.~\eqref{182206_17Oct17}, except for irrelevant additive constants. The binary mixture and the virial descriptions of the RLG are therefore physically equivalent.

\subsection{RLG and the spherical perceptron limit}
\label{sec:perceptron}
From a completely different viewpoint, the RLG can be constructed as a limit case of a continuous satisfaction problem called the perceptron~\cite{rosenblatt1958}. In order to understand this relation, we first recall the definition of the spherical perceptron~\cite{gardner1988optimal,franz2016simplest} in the following.

Consider $N$ points represented by $(d+1)$-dimensional vectors $\by_i\in \mathbb{R}^{d+1}$, $i=1,\dots, N$,
satisfying the spherical constraints
\[
\by_i\cdot\by_i = R^2,\qquad  i = 1,\dots, N.
\]
The the problem is then to find the state vector $\bx\in \mathbb{R}^{d+1}$
that satisfies the exclusion constraints of size $\kappa$
\[
h_i= \bx\cdot\by_i-\kappa\geq 0, \qquad i = 1,\dots, N,
\]
under the spherical constraint $\bx\cdot\bx= R^2$. The partition function of this constraint satisfaction problem (CSP) is given by \begin{align}
Z_{\rm CSP} = 
    \int \dd\bx \delta(\bx\cdot\bx-R^2)\prod_{i=1}^N \theta(\bx\cdot\by_i-\kappa),
\label{eq:zcsp}
\end{align}
which, in the limits $d\to\infty$, $R\to\infty$, and $N\to\infty$ with fixed $R^2/d$ and $N/d$, 
Gardner and Derrida solved for $\kappa\geq 0$~\cite{gardner1988optimal},
and later Franz and Parisi solved for $\kappa< 0$~\cite{franz2016simplest}.

In order to identify the connection between the perceptron and RLG, we consider the RLG on the surface of the $(d+1)$-dimensional hypersphere of radius $R$, instead of its traditional description in $d$-dimensional Euclidean space. The partition function is then
\begin{equation}
\begin{aligned}
    Z_{\rm RLG} &= 
    \int \dd \bx \delta(\bx\cdot\bx-R^2)\prod_{i=1}^N \theta(\abs{\bx-\by_i}-\sigma) \\
    &= \int \dd\bx \delta(\bx\cdot\bx-R^2)\prod_{i=1}^N \theta(\bx\cdot\by_i+R^2-\sigma^2/2).
\label{eq:zrlg}
\end{aligned}
\end{equation}
Interestingly, Eqs.~\eqref{eq:zcsp} and \eqref{eq:zrlg} suggest that the RLG on the hypersphere can be mapped into the spherical version of the perceptron by setting $\kappa = \sigma^2/2-R^2$. 

The RLG in Euclidean space can be identified with the its hypersphere perceptron-like formulation in the thermodynamic limit $R\to\infty$, with fixed $d$ and with the number of obstacles $N$ scaling such that a finite density of obstacles is maintained. (The curvature of the hypersphere is then negligible.) A (naive) expectation might thus be that the solution of the RLG in Euclidean space can be recovered by taking the $\kappa\to -\infty$ limit of the solution of the perceptron derived by Franz and Parisi~\cite{franz2016simplest}. 
This expectation would be valid, provided the limit $d\to\infty$ with fixed $R^2/d$ and $N/d$ is equivalent to the limit $R\to\infty$ with fixed $d$ and finite density of obstacles, followed by $d\to\infty$ with a proper scaling of density. This treatment, however, involves a non-trivial exchange of limits, which sheds some doubt on its validity.

To test out the idea, we write the partition function of the RLG on the hypersphere as
\begin{align}
Z_{\rm RLG} = \int \dd \bx \delta(\bx\cdot\bx-R^2)G(\bx),
\end{align}
where 
\[
G(\bx) = \prod_{i=1}^N \theta(\abs{\bx-\by_i}-\sigma).
\]
We write the delta-function constraint using an integral representation, and  evaluate it via a saddle point method:
\begin{align}
Z_{\rm RLG} \sim \int \dd \bx \int \dd \lambda     e^{-\frac{\lambda}{2}\left(\bx\cdot\bx-R^2\right)}G(\bx) 
    \sim e^{S(\lambda^*)},
\end{align}
where 
\[
S(\lambda) =   \frac{\lambda R^2}{2} + \log\int \dd \bx     e^{-\frac{\lambda}{2}\bx\cdot\bx}G(\bx),
\]
with $\lambda^*$ determined by the (saddle-point) condition
\begin{align}
\left.\pdv{S}{\lambda}\right|_{\lambda=\lambda^*}=0
\to
R^2 = \frac{\int \dd \bx e^{-\frac{\lambda}{2}\bx\cdot\bx} G(\bx)\bx\cdot\bx}{\int \dd \bx e^{-\frac{\lambda}{2}\bx\cdot\bx} G(\bx)}.
 \label{saddle}
\end{align}
Similarly the distribution function of $\bx$ is calculated as 
\begin{align}
\rho(\bx) =
\frac{\delta(\bx\cdot\bx-R^2)G(\bx)}{\int \dd\bx \delta(\bx\cdot\bx-R^2)G(\bx)}
\sim 
\frac{e^{-\frac{\lambda^*}{2}\bx\cdot\bx}G(\bx)}{\int \dd\bx e^{-\frac{\lambda^*}{2}\bx\cdot\bx}G(\bx)}.
\end{align}
If $\lambda^*=0$, then the distribution $\rho(\bx)$ is the same as that of the RLG in Euclidean space. Is this condition satisfied in the thermodynamic limit? To answer this question, we introduce an auxiliary function $f(\lambda^*)$ as
\begin{align}
\ave{\bx\cdot\bx} = d \ave{x_1^2} \equiv d f(\lambda^*),
\end{align}
where we denote
\[
\ave{\bullet} = 
\int \dd\bx \rho(\bx)\bullet = 
\frac{\int \dd \bx e^{-\frac{\lambda^*}{2}\bx\cdot\bx} G(\bx)\bullet}{\int \dd \bx e^{-\frac{\lambda^*}{2}\bx\cdot\bx} G(\bx)},
\]
and we note that $f(\lambda^*)$ is a decreasing function of $\lambda^*$ because
\begin{align}
d\diff{f(\lambda^*)}{\lambda^*} = -\ave{(\bx\cdot\bx)^2}+\ave{\bx\cdot\bx}^2\leq 0.
\end{align}
The equality holds if and only if $\rho(\bx)$ is a delta function. The saddle point condition is now
\begin{align}
f(\lambda^*) = \frac{R^2}{d}.\label{eq:f}
\end{align}
For finite $d$, the right-hand side of Eq.~\eqref{eq:f} diverges in the thermodynamic limit $R\to\infty$.
In this case, one concludes that $\lambda^*\to 0$, because $f(\lambda^*)$ is a decreasing function.
The distribution of the RLG on the hypersphere can thus be identified with that in Euclidean space, as expected. 
However, this does not mean that the solution of the spherical perceptron (See Ref.~\citen{Franz2017}) can be directly used for the RLG in Euclidean space,
because it is derived under the condition that the ratio $R^2/d$ is kept finite, which leads to $\lambda^*>0$. It can be explicitly checked within the exact solution of the perceptron model that $\lambda^*$ remains finite even in the limit $\k\to-\infty$, see Appendix~D in Ref.~\citen{Franz2017}.
We conclude that the limit $\k\to-\infty$ of the perceptron solution does not coincide with the solution of the RLG in Euclidean space in~$d\to\infty$. 

\section{Dynamical Derivation}
\label{sec:dynamic}
\label{sec:dyncavity}

We now turn to the $d\to\infty$ dynamics of the RLG. Using a dynamical cavity treatment, we here show that the static analogy between the RLG and a hard sphere liquid in the limit $d\rightarrow\infty$ also holds for the dynamics.
Note that the notation used here follows that of Agoritsas \textit{et al.}~\cite{agoritsas2018out}, which differs slightly from that of Ref.~\citen{maimbourg2016solution}. More specifically, $m$ is the tracer's mass, $\zeta$ is the friction coefficient of an isolated tracer, $\boldsymbol{\xi}(t)$ is the Gaussian white noise acting on the tracer with auto-correlation
function $\langle\boldsymbol{\xi}(t) \boldsymbol{\xi}(t')\rangle = 
2\zeta \boldsymbol{I} \delta(t-t')$. Note also that, as discussed by Manacorda \emph{et al}.~\cite{manacorda2020numerical}, a convenient way to analyze the motion of the tracer among hard obstacles is to consider first its motion among purely repulsive yet softened obstacles, described by a continuous dimensionless potential $V(r)$, and to take the hard sphere limit at the end of the computation. This approach sidesteps technical difficulties associated with the singular nature of the hard sphere potential, but without affecting the final result. To simplify the notation, and without loss of generality, we also set the inverse temperature to unity, $\beta=1$. 

The standard dynamical cavity approach consists of:
\begin{enumerate}
\item writing equations of motion for the original system;
\item adding an additional variable to the problem, 
in such a way that its effect on the original system is small;
\item treating this addition perturbatively to derive an equation of motion 
for the evolution of the new variable;
\item obtaining a self-consistent equation for a \emph{typical}
variable by noting that the new variable is identical to all existing variables
in the system.
\end{enumerate}
More specifically, in our implementation, the second step consists of changing the dimension of the system, $d\to d+1$. This choice is inspired by Agoritsas 
\textit{et al.}'s analysis of the perceptron, in which 
the dimension of the problem, \textit{i.e.}, the dimension of the 
perceptron sphere, is similarly increased.  
Note that the presentation below is intended to be physically intuitive. Careful order-of-magnitude estimates of the various terms can be found in Ref.~\citen{Chen2020}, which presents a related mean-field approach for a variety of similar problems. Note also that (as in the perceptron model) there are three sources of randomness: the initial condition, the thermal noise, and the quenched disorder. Averaging over the initial condition and noise is here denoted as $\langle \cdots \rangle$, and averaging over disorder as $\overline{\cdots}$. These averages initially pertain to the original
$d$-dimensional system, as indicated by the subscript to the averaging notation.
Additional averages are defined later, as needed.

Let $x_\mu(t)$ be the tracer coordinates, which satisfy the equation of motion,
\begin{eqnarray}\label{unpert}
\zeta \partial_t x_\mu(t) = -\partial_{x_\mu(t)} \sum_i V(|\bx(t)-\by_i|)
+ \xi_\mu(t)
\end{eqnarray}
for $\mu=1, ..., d$, where $\bx\equiv \left(x_1, ..., x_d\right)$ is the $d$-dimensional vector,
$\by_i \equiv \left(R_{1,i}, ..., R_{d,i} \right)$ is the position of the $i$th obstacle in $d$-dimensional space, and $\xi_\mu(t)$ is the $\mu$th component of the $d$-dimensional Gaussian white noise $\boldsymbol{\xi}(t)$. We here explicitly discuss only the overdamped case, which corresponds to vanishing mass, $m\to 0$, but the derivation is completely general. It also applies to underdamped (Langevin) dynamics, and the inertial term involving the acceleration could be restored following Ref.~\citen{agoritsas2018out}.

The initial tracer position, $x_\mu$, is chosen from the equilibrium Gibbs probability distribution, 
\begin{eqnarray}\label{unpertpdf}
P_d[\bx] = \frac{1}{Z_{d,N}} 
\exp\left[-\sum_i V(|\bx-\by_i|)\right],
\end{eqnarray}
where $Z_{d,N}$ is the partition function for a system with dimension $d$ and $N$ obstacles specified explicitly. In the hard sphere limit distribution Eq.~\eqref{unpertpdf} becomes a uniform distribution in the void space and 
the partition function corresponds to the volume of void space left by obstacles, which are quenched independent random variables. 

Following the above scheme, we now increase the spatial dimension of the problem, which entails adding a new component to the tracer position $x_0(t)$, a new noise component, $\xi_0(t)$, as well as an additional component to the vectors specifying obstacle locations, $R_{0,i}$, $i=1, ..., N$.
In the presence of this additional variable the equation of motion for the original variables (Eq.~\eqref{unpert}) becomes
\begin{equation}\label{pert1}
\zeta \partial_t x_\mu(t) = -\partial_{x_\mu(t)} \sum_i 
V_i(\bx(t)) + \xi_\mu(t), \hskip 1em \mu=1, ..., d,
\end{equation}
where we use the shorthand 
\[
V_i(\bx(t)) \equiv V [ ( |\bx(t)-\by_i|^2+|x_0(t)-R_{0,i}|^2 )^{1/2} ]
\]
in the following. 

To leading order in $|x_0-R_{0,i}|/|\bx -\by_i|$, which we implicitly treat as an $\mathcal{O}(1/\sqrt{d})$ term,
the perturbed equation of motion for the original variables is then
\begin{eqnarray}\label{pert3}
\zeta \partial_t x_\mu(t) =&&
\nonumber  
- \partial_{x_{\mu}(t)} \sum_i \Big[ V(|\bx(t) -\by_i|) 
\nonumber \\ && 
+ \frac{V'(|\bx(t) -\by_i|)}{|\bx(t)-\by_i|}
h_{0i}(t)
\Big] + \xi_\mu(t), 
\end{eqnarray} 
where $V'(r)=\dd V(r)/\dd r$ and $h_{0i}(t)  = \frac{1}{2}|x_0(t)-R_{0,i}|^2$. 
Note that $h_{0i}$ does not have a simple geometric interpretation. In particular, it is \emph{not} a change of the gap between the particle and obstacle $i$. It nevertheless does not depend on $\bx$ and $\by_i$, which simplifies some of the following considerations.

The initial condition for $x_\mu$ in Eq.~\eqref{pert3} is chosen from the 
equilibrium Gibbs probability distribution, in which the additional dimension acts as an external field,
\begin{eqnarray}\label{pertpdf}
&& P_d(\bx| x_0) \approx \frac{1}{Z_{d,N}} 
\exp\left\{-\sum_i V(|\bx-\by_i|)
\right. \nonumber \\ && \left. - \sum_i
\left[\frac{V'(|\bx-\by_i|)}{|\bx-\by_i|} 
- \left<\frac{V'(|\bx-\by_i|)}{|\bx-\by_i|}
\right>_d \right] h_{0i} \right\} 
\end{eqnarray}
Following Agoritsas \textit{et al.}~\cite{agoritsas2018out}, we can write the trajectories of the
original variables as the sum of the unperturbed trajectories and of the trajectories perturbed by 
the change in both the equation of motion and the initial condition. We then obtain
\begin{equation}\label{pert4}
x_\mu(t) = x_\mu^{(0)}(t) + x_\mu^{\text{(dyn)}}(t) + x_\mu^{\text{(in)}}(t),
\end{equation}
where $x_\mu^{(0)}(t)$ satisfies the unperturbed equation of motion
with an initial condition drawn from the unperturbed ensemble given by Eq.~\eqref{unpertpdf},
$x_\mu^{\text{(dyn)}}(t)$ is the trajectory change originating from the second term in the perturbed equation of motion given by Eq.~\eqref{pert3}, and $x_\mu^{\text{(in)}}(t)$ is the trajectory change 
originating from the perturbed initial condition given by Eq.~\eqref{pertpdf}.
Agoritsas \textit{et al.}~\cite{agoritsas2018out} formally wrote the latter two components as
\begin{subequations}
\begin{align}
\label{pert5}
x_\mu^{\text{(dyn)}}(t) &= \sum_i \int_0^t \dd t' 
\frac{\delta x_\mu^{(0)}(t)}{\delta h_{0i}(t')} h_{0i}(t'),
\\ \label{pert6}
x_\mu^{\text{(in)}}(t) &= \sum_i \frac{\delta x_\mu^{(0)}(t)}{\delta h_{0i}}
h_{0i},
\end{align}
\end{subequations}
where in Eq.~\eqref{pert6} $h_{0i}\equiv h_{0i}(t=0)$

Let us now consider the equation of motion for the new coordinate of the tracer,
\begin{equation*} 
\zeta \partial_t x_0(t) = - \partial_{x_0(t)} \sum_i V_i(\bx(t)) + \xi_0(t).
\end{equation*}
Using Eqs.~\eqref{pert4}-\eqref{pert6} we can write 
\begin{equation}\label{newv2} 
\begin{aligned}
\zeta \partial_t x_0(t) \approx 
&-\partial_{x_0(t)} \sum_i V_i(\bx^{(0)}(t)) \\
&-\partial_{x_0(t)} \sum_{i,j} \frac{ \delta V_i(\bx^{(0)}(t)) } {\delta h_{0j}}h_{0j} \\
&-\partial_{x_0(t)} \sum_{i,j} \int_0^t \dd t' 
\frac{ \delta  V_i(\bx^{(0)}(t))} {\delta h_{0j}(t')} h_{0j}(t') + \xi_0(t),
\end{aligned}
\end{equation} 
where we denote again the shorthand 
\[
V_i(\bx^{(0)}(t)) \equiv V[(|\bx^{(0)}(t)-\by_i|^2+|x_0(t)-R_{0,i}|^2)^{1/2}].
\]
The first term on the right-hand side (RHS) of this equation is a fluctuating potential field at position $x_0(t)$. Its fluctuations are
due to the unperturbed ``gap'' variables, 
$r^{(0)i}(t) =|\bx^{(0)}(t)-\by_i|$,
which evolve on their own and due to the quenched randomness, $R_{0,i}$. 
The second and third terms describe a feedback process. The
additional coordinate perturbs the tracer evolution in the original $d$-dimensional system, which 
in turn influences its evolution in the additional dimension. 

In order to complete the derivation, we make two assumptions whose justification will be presented elsewhere~\citen{Chen2020}:
\begin{enumerate}
\item The influence of the presence of a specific obstacle on the distance between the tracer and another specific obstacle is negligible. This implies that the contributions to the force originating from different obstacles are uncorrelated and therefore only the diagonal $i=j$ terms contribute to the double summations in Eq.~\eqref{newv2}. 
\item The summations in the second and third terms of Eq.~\eqref{newv2} concentrate around their averages. Note that these averages include averaging over the disorder, including the $0$th coordinates of the obstacles, \textit{i.e.} disorder averages are $d+1$ dimensional.  
\end{enumerate}
The second assumption leads to the following 
\begin{subequations}
\begin{align}\label{selfave1}
\sum_{i} \frac{ \delta V_i(\bx^{(0)}(t)) } {\delta h_{0i}(t')} h_{0i}(t') &\to 
\sum_i \subline{\frac{\delta \left< V_i(\bx^{(0)}(t)) \right>_d }
{\delta h_{0i}(t')}h_{0i}(t')}{d+1},
\\
\label{selfave2}
\sum_i \frac{ \delta V_i(\bx^{(0)}(t))  } {\delta h_{0i}} h_{0i} &\to  
\sum_i \subline{\frac{ \delta \left< V_i(\bx^{(0)}(t)) \right>_d^h }
{\delta h_{0i}}h_{0i}}{d+1},
\end{align}
\end{subequations}
where $\left< \dots \right>_d^h$ denotes averaging over distribution of Eq.~\ref{pertpdf}
and the functional derivatives are evaluated at $h_{0i}= 0$.

The two functional derivatives above are related through the fluctuation-dissipation relation. To introduce this relation we first note due to Eq.~\eqref{pertpdf} the functional 
derivative in Eq.~\eqref{selfave2} can be expressed in terms of an equilibrium time-dependent correlation function.
For a given obstacle $i$ we then get (recalling that $\beta=1$)
\begin{eqnarray}\label{FDTa}
\frac{ \delta \langle 
V_i(\bx^{(0)}(t)) \rangle_d^h}
{\delta h_{0i}} = - &\left\langle 
V_i(\bx^{(0)}(t))
\left[\frac{V'(|\bx^{(0)}-\by_i|)}{|\bx^{(0)}-\by_i|}  
\right.\right. \new &\quad  \left.\left.
 - \left\langle\frac{V'(|\bx^{(0)}-\by_i)}
{|\bx^{(0)}-\by_i|}\right\rangle_d \right]
\right>_d,
\end{eqnarray}
where $\bx^{(0)}\equiv\bx^{(0)}(t=0)$.

More generally, let us define the following correlation function
\begin{eqnarray}\label{Adef}
A_i(t,t') = - &\left\langle V_i(\bx^{(0)}(t))  \left[\frac{V'(|\bx^{(0)}(t')-\by_i|)}
{|\bx^{(0)}(t')-\by_i|} 
\right.\right. \new &\quad \left.\left.
- \left<\frac{V'(|\bx^{(0)}(t')-\by_i|)}
{|\bx^{(0)}(t')-\by_i|}
\right>_d \right] \right\rangle_d.
\end{eqnarray}
In terms of this new correlation function, the fluctuation-dissipation relation reads, 
\begin{eqnarray}\label{FDT}
 \frac{ \delta \left\langle 
V_i(\bx^{(0)}(t)) \right\rangle_d}
{\delta h_{0i}(t')} =
- &\partial_{t'}
\left\langle V_i(\bx^{(0)}(t)) \left[\frac{V'(|\bx^{(0)}(t')-\by_i|)}
{|\bx^{(0)}(t')-\by_i|} 
\right.\right. \new & \left.\left.
- \left<\frac{V'(|\bx^{(0)}(t')-\by_i|)}
{|\bx^{(0)}(t')-\by_i|}
\right>_d \right] \right\rangle_d.
\end{eqnarray}
\begin{widetext}
Using the fluctuation-dissipation relation and integrating by parts we obtain
\begin{equation}\label{newv3}
\begin{aligned}
\zeta \partial_t x_0(t) =
&- \partial_{x_0(t)}
\sum_i V_i(\bx^{(0)}(t)) 
- \partial_{x_0(t)} \sum_i \subline{A_i(t,t)
h_{0i}(t)}{d+1} \\ 
&- \partial_{x_0(t)} \int_0^t \dd t' 
\sum_i 
\subline{
\left<V_i(\bx^{(0)}(t)) 
\left(\frac{V'(|\bx^{(0)}(t')-\by_i|)}
{|\bx^{(0)}(t')-\by_i|} 
- \left<\frac{V'(|\bx^{(0)}(t')-\by_i|)}
{|\bx^{(0)}(t')-\by_i|}
\right>_d \right) \right>_d \partial_{t'} h_{0\nu}(t')}{d+1} \\ 
&+ \xi_0(t)
\end{aligned}
\end{equation}
It follows from the original equation of motion
for the additional coordinate that the differentiation $\partial_{x_0(t)}$ acts only on $x_0(t)$ in $V_i(\bx^{(0)}(t))$. 

Because every quantity except $x_0(t)$ is
integrated out, by translational symmetry the expression 
$\sum_i \subline{A_i(t,t)h_{0i}(t)}{d+1}$
in the second term of Eq.~\eqref{newv3} is a constant and hence its contribution vanishes. The subtracted contribution in the penultimate term vanishes as well, because it involves averaging over the tracer position, in the presence of one specific obstacle. In the overwhelming majority of phase space, the tracer is far from this obstacle. Even when the tracer is localized, it spends most of its time localized away from that obstacle. The penultimate term can thus be rewritten as
\begin{eqnarray}\label{meaning2}
& &- \sum_i \subline{ \partial_{x_0(t)} 
\left<V_i(\bx^{(0)}(t))
\frac{V'(|\bx^{(0)}(t')-\by_i|)}
{|\bx^{(0)}(t')-\by_i|} \right>_d \partial_{t'} h_{0\nu}(t')}{d+1}
\nonumber \\ &=&
- \sum_i \subline{ \partial_{x_0(t)} 
\left<V_i(\bx^{(0)}(t))
\frac{V'(|\bx^{(0)}(t')-\by_i|)}
{|\bx^{(0)}(t')-\by_i|} \right>_d (x_0(t')-R_{0,i})}{d+1}
\partial_{t'}x_0(t'),
\end{eqnarray}
\end{widetext}
where the average can be identified with the auto-correlation
function of the force acting along the $0$th coordinate of the tracer, 
\begin{equation}\label{meaning3} \begin{aligned}
F_{0,i}(t) &= - V'(|\bx^{(0)}(t)-\by_i|)
\frac{x_0(t)-R_{0,i}}{|\bx^{(0)}(t)-\by_i|}  \\
&\approx - \partial_{x_0(t)}V_i(\bx^{(0)}(t)).
\end{aligned} \end{equation}
The average is therefore given by
\begin{eqnarray}\label{meaning5}
\sum_i \subline{\left<F_{0,i}(t) F_{0,i}(t')\right>_d}{d+1},
\end{eqnarray}
which depends on the displacement of the tracer along the additional
coordinate, $x_0(t)-x_0(t')$. We note that during the decay of the auto-correlation function in Eq.~\eqref{meaning5} the tracer displacement along the $0$th coordinate is vanishingly small and therefore, to leading order in $1/d$, we can average Eq.~\eqref{meaning5} over all possible values of $x_0(t)-x_0(t')$, which results in
\begin{eqnarray}\label{meaning6}
\sum_i \subline{\left<F_{0,i}(t) F_{0,i}(t')\right>_{d+1}}{d+1}.
\end{eqnarray}
(A more careful discussion of this point will be given in Ref.~\citen{Chen2020}.) 

The equation of motion for the additional coordinate is then
\begin{eqnarray}\label{newv5}
&& \zeta \partial_t x_0(t) =
\sum_{i} F_{0,i}(t) 
\\ \nonumber && - \sum_{i} \int_0^t \dd t'  
\subline{\left<F_{0,i}(t) F_{0,i}(t')\right>_{d+1}}{d+1}
\partial_{t'}x_0(t') 
+ \xi_0(t).
\end{eqnarray}
In the $d\rightarrow\infty$ limit, this coordinate is indistinguishable from any other coordinate. We can then write the memory function as
\begin{equation}\label{memory1} \begin{aligned}
& &\sum_{i} \subline{\left<F_{0,i}(t) F_{0,i}(t')\right>_{d+1}}{d+1} \\
&=& \frac{1}{d} \sum_{i} \subline{\left<\mathbf{F}_i(t) \cdot
\mathbf{F}_i(t')\right>_{d+1}}{d+1}.
\end{aligned} \end{equation}
Next, in this limit the random force, \textit{i.e.} the first term on the RHS of Eq.~\eqref{newv5} becomes a Gaussian random variable $\eta_0$. Finally, recognizing once more that in the $d\rightarrow\infty$ limit all coordinates are equivalent we can write the equation of motion for the tracer in vector notation,
\begin{eqnarray}\label{newv6}
\zeta \partial_t \mathbf{x}(t) &=&    
\boldsymbol{\eta}(t) - \int_0^t \dd t' M(t-t')\partial_{t'}\mathbf{x}(t')
+ \boldsymbol{\xi}(t),
\end{eqnarray}
\begin{eqnarray}\label{newv7}
\langle\boldsymbol{\eta}(t)\boldsymbol{\eta}(t')\rangle = 
\boldsymbol{I} M(t-t').
\end{eqnarray}

The memory function in Eq.~\eqref{memory1} depends on the statistics of
$r_{\mu,i}(t) \equiv x_\mu(t)-R_{\mu,i}$ only. 
To complete the derivation we need to repeat the above
treatment for $r_{0,1}(t)\equiv x_0(t)-R_{0,1}$ and
once again recognize the equivalence of all coordinates. 
We start with the equation of motion for $r_{0,1}$,
\begin{equation}\label{newvd1} \begin{aligned}
\zeta \partial_t r_{0,1}(t) =
&- \partial_{r_{0,1}(t)} V\left( \sqrt{|\mathbf{r}_1(t)|^2+(r_{0,1}(t))^2 }\right) \\
&- \sum_{i\neq 1} \partial_{r_{0,i}(t)} 
V\left( \sqrt{ |\mathbf{r}_i(t)|^2+(r_{0,i}(t))^2}\right) \\
&+ \xi_0(t).
\end{aligned} \end{equation}
The second term on the RHS can be analyzed as before.
Because the tracer interacts with an average of $d$ obstacles, in the limit $d\rightarrow\infty$ excluding one specific
obstacle does not matter. The resulting equation of motion is
\begin{equation}\label{newvd2} \begin{aligned}
\zeta \partial_t r_{0,1}(t) 
&= F_{0,1}(t) +\sum_{i\neq 1} F_{0,i}(t) \\ 
&- \sum_{i\neq 1} \int_0^t \dd t' \subline{\left<F_{0,i}(t) F_{0,i}(t')\right>_d}{d+1} \partial_{t'}r_{0,1}(t') \\ 
&+ \xi_0(t),
\end{aligned} \end{equation}
where 
\begin{eqnarray}\label{meaning8}
F_{0,i}(t) = - \partial_{r_{0,i}(t)} 
V\left(\sqrt{|\mathbf{r}^{(0)}_i(t)|^2+(r_{0,i}(t))^2}\right).
\end{eqnarray}
Again, replacing Eq.~\eqref{meaning5} by Eq.~\eqref{meaning6}
gives the standard expression,
\begin{equation}\label{newvd3} \begin{aligned}
&\zeta \partial_t r_{0,1}(t) = F_{0,1}(t) 
+\sum_{i\neq 1} F_{0,i}(t) \\ 
&- \sum_{i\neq 1} \int_0^t \dd t' 
\subline{\langle F_{0,i}(t) F_{0,i}(t')\rangle_{d+1}}{d+1}
\partial_{t'}r_{0,1}(t') 
+ \xi_0(t).
\end{aligned} \end{equation}

The central argument of the cavity method can then be invoked. In the limit $d\rightarrow\infty$, 
$r_{0,1}(t)$ is equivalent to any $r_{\mu,1}(t)$, 
$\sum_{i\neq 1} F_0^i(t)$ becomes a random Gaussian force with autocorrelation,
$\subline{\left<F_0^i(t) F_0^i(t')\right>_{d+1}}{d+1}$, self-consistently
determined by the statistics of $r_{\mu,i}(t) = x_\mu(t)-R_{\mu,i}$, 
which are the same as for $r_{\mu,1}(t)$, \textit{etc}.
We thus have the self-consistent stochastic process,
\begin{eqnarray}\label{newvd4}
\zeta \partial_t \mathbf{r}_1(t) &=& \mathbf{F}(\mathbf{r}_1(t))  
+ \boldsymbol{\eta}(t) \nonumber\\
& &- \int_0^t \dd t' M(t-t')\partial_{t'}\mathbf{r}_1(t') 
+ \boldsymbol{\xi}(t),
\end{eqnarray}
where, consistently with Eq.~\eqref{newv7},
\begin{eqnarray}\label{rndmforce}
\left<\boldsymbol{\eta}(t)\boldsymbol{\eta}(t')\right> = 
\boldsymbol{I} M(t-t')
\end{eqnarray}
but with an explicit expression for the memory function,
\begin{eqnarray}\label{mfnction}
M(t) = \frac{n}{d} \int \dd \mathbf{r}_1 e^{-V(r_1)}
\left<\mathbf{F}(\mathbf{r}_1(t))\cdot\mathbf{F}(\mathbf{r}_1)\right>_{\mathbf{r}_1},
\end{eqnarray}
where in turn the average $\langle \dots \rangle_{\mathbf{r}_1}$ is over the stochastic process defined in Eq.~\eqref{newvd4}.

Equations \eqref{newvd4}-\eqref{mfnction} allow one to evaluate the
memory function, which can then be used to 
analyze the motion of the tracer from Eqs.~\eqref{newv6}-\eqref{newv7}.
One can then follow the derivation in one of the recent references on the topic~\cite{szamel2017simple,agoritsas2018out,francesco2020theory}, and derive the self-consistent equation for the localization length from Eqs.~\eqref{newvd4}-\eqref{mfnction}. The condition of Eq.~\eqref{154732_15Feb18}, which gives that a discontinuous dynamical transition arrests the tracer dynamics in the limit $d\rightarrow\infty$, is straightforwardly recovered. In the $d\rightarrow\infty$ limit, the dynamics of the RLG is therefore completely equivalent to that of an equilibrium hard sphere liquid, after a mere factor of two rescaling.

\section{Numerical results and discussion}
\label{sec:numerical}

The above theoretical approaches provide a consistent mean-field, $d\rightarrow\infty$ description of RLG caging. However, they all generically leave out perturbative corrections of leading order $1/d$, let along non-perturbative corrections. As discussed in the context of the static derivations above, the direct cavity reconstruction scheme does not readily provide an estimate of the prefactor for this correction (and neither does the dynamical scheme of dynamic derivation), but a finite-$d$ Gaussian cage ansatz can straightforwardly be implemented in the virial treatment. Although this  ansatz was found not to hold in finite-$d$ glasses in the vicinity of the dynamical arrest~\cite{charbonneau2012dimensional}, and clearly fails to account for non-perturbative hopping corrections near the mean-field dynamical transition at $\hat\varphi_\mathrm{d}$~\cite{biroli2020unifying}, it is unclear whether similar problems affect RLG caging at high obstacle densities. In this regime, hopping is indeed strongly suppressed and the Gaussian ansatz might well predict the scale of perturbative corrections. In this section, we use finite-$d$ simulations to provide an overall evaluation of the Gaussian ansatz.

\begin{figure}[ht]
\centering
\includegraphics[width=0.48\textwidth]{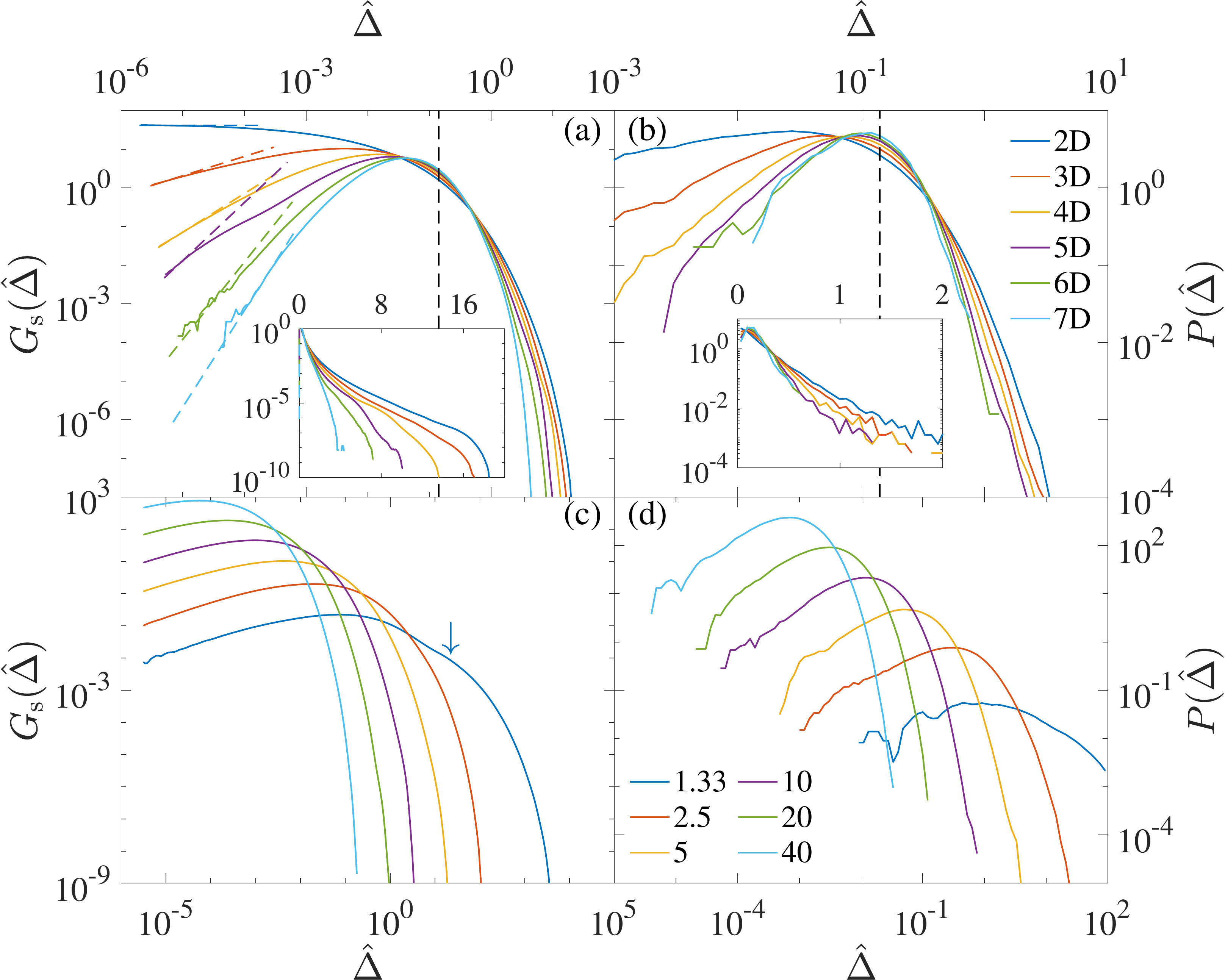}
\caption{(a) Self van Hove function and (b) cage size distribution at $\hat\varphi=5$ for various $d$. The small displacement $G_\mathrm{s}(\hat\Delta) \sim \hat\Delta^{d/2-1}$ (colored dashed lines) and the $d\to\infty$ prediction from Eq.~\ref{eq:cagesizegaussian} (vertical black dashed line) are included. Insets in (a,b): Same results given on a lin-log scale. (c) Self van Hove function and (d) cage size distribution in $d=3$ for various $\hat\varphi$. At small obstacle density, close to the percolation threshold, a shoulder emerges in $G_\mathrm{s}(\hat\Delta)$ (arrow in (c)). This feature is associated with a fat tail that makes the mean cage size $\hat\Delta$ diverge for $\hat\varphi<\hat\varphi_\mathrm{p}$. The cage size distribution also broadens upon decreasing $\hat\varphi$.}
\label{fig:csizedist}
\end{figure}

\begin{figure*}[t]
  \centering
  \includegraphics[width=1\textwidth]{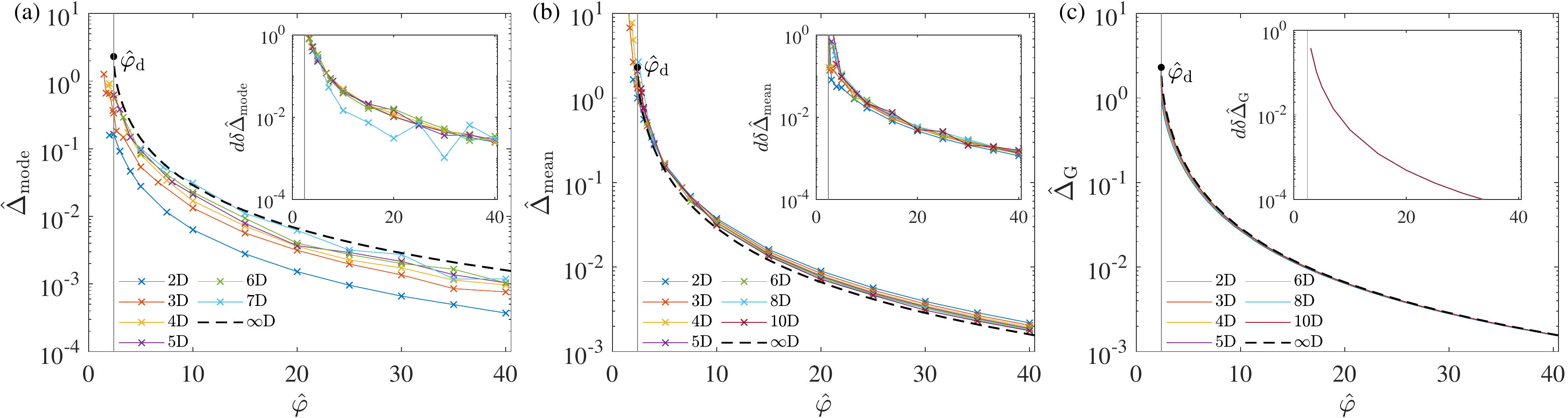}
  \caption{Comparison between various cage sizes: (a) modal cage size extracted from the mode of $P(\hat\Delta)$; (b) mean cage size reported in Ref.~\citen{biroli2020unifying}; (c) Gaussian-ansatz cage size from  Eq.~\eqref{eq:cagesizegaussian}. All three quantities converge to the $d \rightarrow \infty$ result from Eq.~\eqref{154732_15Feb18}   (black dashed lines). The insets show that in all three cases deviations from the $d\to\infty$ results scale as $1/d$, albeit with maarkedly different prefactors. The mode and the mean have different signs and are both at least an order of magnitude larger than the Gaussian-ansatz prediction. }
    \label{fig:csizemetrics}
\end{figure*}

The challenge of assessing the Gaussian ansatz is that it implicitly assumes that all cages are the same size. Although this \emph{typical} cage does dominate in the $d\rightarrow\infty$ limit, strong deviations from caging uniformity are observed in finite $d$. More quantitatively, the mean cage size is obtained by averaging over cages (i.e., disorder) and over initial tracer positions within a cage
\begin{equation}
\hat\Delta = \frac{d}{\mathbb{C}} \sum_\mathbb{C} \frac{1}{|\{i\}||\{j\}|} \sum_{i,j\in\mathbb{C}} (\vec{r}_i - \vec{r}_j)^2,
\end{equation}
where $\mathbb{C}$ is a cage realization. The Gaussian ansatz, however, only accounts for the latter averaging and replaces the disorder average by a concentration assumption. 
Two distribution naturally assess these two effects, and thus allow us to discern the dominant contribution.
\begin{enumerate}
\item The long-time limit of the self part of the van Hove function
\begin{equation} \label{eq:dis_svh}
G_\mathrm{s}(r) \sim \sum_{\mathbb{C},i,j} \delta(|\vec{r}_i - \vec{r}_j|-r),
\end{equation}
normalized as $\int_0^\infty G_\mathrm{s}(r) \dd r = 1$, can be expressed in terms of the squared displacement $\hat\Delta = r^2 d$ through a change of variables, $G_\mathrm{s}(\hat\Delta) = G_\mathrm{s}(r)/(2r)$. The resulting distribution accounts for both types of averaging.
\item The cage size distribution
\begin{equation} \label{eq:dis_csize}
P(\hat\Delta) \sim \sum_{\mathbb{C}} \delta \left( \frac{1}{|\{i\}||\{j\}|} \sum_{i,j} (\vec{r}_i - \vec{r}_j)^2-\hat\Delta/d \right),
\end{equation}
normalized as $\int_0^\infty P(\hat\Delta) \dd \hat\Delta = 1$, accounts for the disorder distribution only.
\end{enumerate} 
Numerically, these observables can be efficiently obtained by cavity reconstruction and Monte Carlo integration as in Refs.~\citen{biroli2020unifying,charbonneau2020percolation}. 
Sample results averaged over $5 \times 10^4$ (in $d=2,3$) to $300$ (in $d=7$) cavities are given in Fig.~\ref{fig:csizedist}.

As expected, both distributions narrow as $d$ increases, and seemingly converge to a $\delta$ distribution as $d\rightarrow\infty$ (Fig.~\ref{fig:csizedist}(a, b)). Conversely, upon decreasing $d$ distributions not only broaden but also become increasingly asymmetric. For instance, the small $\hat\Delta$ limit of the van Hove function scales as $G_\mathrm{s}(r) \sim r^{d-1}$ and thus $G_\mathrm{s}(\hat\Delta) \sim \hat\Delta^{(d-1)/2}$, while in the large $\hat\Delta$ limit, $G_\mathrm{s}(\hat\Delta)$ decays quickly. The cage size distribution $P(\hat\Delta)$ is also lopsided, although slightly less. From this comparison, we conclude that the non-Gaussian character of caging is significantly affected both by cage-to-cage fluctuations and by the non-Gaussian character of individual cages, even at high densities.

Upon approaching $\hat\varphi_\mathrm{p}$ the static cage size distribution further broadens at large $\hat\Delta$ (Fig.~\ref{fig:csizedist}(c, d)), as in the MK model~\cite{charbonneau2014hopping}. This deviation, which is clearly distinct from the non-Gaussian caging correction, contributes to the dramatic growth of the mean cage size in that regime. It also eventually leads to void percolation. The Gaussian ansatz, which assumes that cages are local and closed, thus completely fails in this regime.

Perturbative $1/d$ corrections to the mean cage size, $\hat\Delta$, arise both from the distribution mode shifting with $d$ and from its increasing anisotropy. It is therefore strongly affected by the percolation physics as well as other caging anisotropy, which are bound to lead to deviations from the Gaussian ansatz.
In order to minimize this contribution, we also consider the modal cage size~\cite{biroli2020unifying}, 
\begin{equation}
\hat{\Delta}_\mathrm{mode} = \arg\max (\hat{\Delta} \cdot P( \hat{\Delta} )),
\end{equation}
where $P( \Delta)$ is the probability of having a cage of size $\Delta$ (as shown in Fig.~\ref{fig:csizedist}(b) for example), with $\int_0^\infty \hat{\Delta} P(\hat{\Delta}) \dd \hat{\Delta} = \hat{\Delta}_\mathrm{mean}$. As argued in Ref.~\citen{biroli2020unifying}, this quantity is closer in spirit to the saddle-point evaluation in a mean-field calculation and might thus be a better finite-$d$ estimator of local caging. 
For each of these different caging estimators (est), we define the deviation from the $d\rightarrow\infty$ result in Eq.~\eqref{eq:csizedeviation} as
\begin{equation} \label{eq:csizedeviation}
\delta \hat\Delta_\mathrm{est} = |\hat\Delta_\mathrm{est} - \hat\Delta_{d \rightarrow \infty}| \sim (d \hat\Delta_\mathrm{est}) / d,
\end{equation}
where $d \hat\Delta_\mathrm{est}$ is the prefactor of the perturbative correction.

Figure~\ref{fig:csizemetrics} shows the evolution of different cage size estimators with packing fraction $\hat\varphi$. As expected, both the modal and the mean cage sizes converge nicely to $\hat{\Delta}_\mathrm{d\rightarrow\infty}$ at densities far above the percolation threshold, and only the former exhibit a perturbative scaling in the surroundings of $\hat\varphi_\mathrm{p}$. The contribution of percolation physics can thus be reasonably well isolated.
The deviation prefactors of these two estimators, however, have different signs. They are also at least an order of magnitude larger (in absolute value) than the Gaussian-ansatz correction from Eq.~\eqref{eq:qagaussian}. This mismatch reflects the pronounced anisotropy of the van Hove and cage size distributions, an effect that is completely absent in the Gaussian ansatz. In addition, as density approaches $\hat\varphi_\mathrm{d}$, the modal correction appears to diverge, while the Gaussian ansatz remains finite. This discrepancy is not a mere quantitative concern, but qualitatively wrong. This effect is likely related to the expectation that the dynamical susceptibility should diverge around $\hat\varphi_\mathrm{d}$ in the $d\to\infty$ limit. Hence, the Gaussian ansatz not only misses out on percolation physics, but seems to omit key features of mean-field physics as well.

\section{Conclusion}
\label{sec:conclusion}
This article contains two main sets of results. The first is methodological: we develop several complementary approaches to analyse 
the RLG in the $d\rightarrow\infty$ limit. We show that the static approach based on replica theory, for which we present two different derivations, agrees with a full dynamical treatment obtained through the cavity method. 
However, we also unveil that despite the apparent similarity between perceptron dynamics and RLG, the mean-field solution of the perceptron does not allow to recover the RLG in the $d\rightarrow\infty$ limit.
The second main outcome of our work has been obtained by contrasting results of numerical simulations in very high dimension with the $d\rightarrow\infty$ analytical solution of the RLG. This comparison allowed us to confirm the distinct role of percolation and glassy physics in the $d\rightarrow\infty$ limit and to clarify to what extent the mean-field solution is able to capture the latter but not the former. Our work highlights that corrections to the mean-field solution are crucial to describe dynamics even in large but finite dimension and opens the way for their analysis. 
In fact, we have found that cage formation and caging dynamics  are well described by the mean-field solution for 
$d \rightarrow \infty$, and that finite-dimensional corrections can be captured perturbatively but require  analytical treatments beyond the Gaussian-ansatz employed until now. 
Cage escape, which contributes to percolation, is a more subtle phenomenon that takes place on timescales that seemingly diverge exponentially with $d$. 
It thus corresponds to a non-perturbative instantonic correction~\cite{biroli2020unifying}, which may only be captured by large-deviation calculations (in contrast to the saddle point calculations presented in the previous sections).
In conclusion, the RLG -- despite its simplicity and its differences with models of supercooled liquids -- offers a new way to look at the problem of glass transition and provides the long-sought framework and guideline to tackle some of the important perturbative and non-perturbative effects in glassy dynamics.  

The dynamical solution of the RLG is also particularly interesting in the context of the mode-coupling theory (MCT) of the same system. MCT has long been understood as a mean-field theory of the glass and localization transitions. This analogy would imply that it should fare particularly well in the $d\rightarrow\infty$ limit, where a proper mean-field theory becomes exact. In the context of the glass transition, however, this expectation has been shown to be only partially true. MCT correctly predicts the critical features of the infinite dimensional dynamic glass transition (\textit{i.e.} its discontinuous nature) but fails rather spectacularly at predicting the location of this transition~\cite{ikeda2010mode,schmid2010glass,Bouchaud2010mct,charbonneau2011glass,maimbourg2016solution}. For the RLG the situation is a bit more involved. The high-dimensional limit of the full wavevector-dependent MCT predicts a discontinuous localization transition~\cite{jin2015dimensional}, which qualitatively agrees with the exact dynamical transition. As for the dynamical glass transition, MCT predicts an incorrect $d\rightarrow\infty$ scaling of the localization transition. However, in finite dimensions MCT predicts a continuous localization transition, which agrees with numerical simulations. In contrast, as we have shown here, the exact infinite dimensional dynamic theory, when generalized, perhaps too naively, to finite dimensions, predicts a discontinuous localization transition. It would be very interesting to look for a less naive generalization of the infinite dimensional theory that correctly predicts the character of the finite dimensional localization transition. Conversely, it would also be instructive to reformulate MCT to properly capture the scaling of the localization threshold, its mean-field criticality as well as the rich interplay between continuous and discontinuous caging at high yet not diverging $d$. 

\begin{acknowledgements}
We thank Dave Thirumalai for inviting us to contribute to this special issue in his honor. We also want to acknowledge the key role of his (and his collaborators’) ideas on our own research path. We also thank all our own collaborators on these topics for their numerous and invaluable inputs.
In particular, we acknowledge many stimulating discussions with E.~I.~Corwin and with B.~Charbonneau. This work was supported by the Simons Foundation grants \#454937 (PC), \#454935 (BG), and \#454955 (FZ). The computations were carried out on the Duke Compute Cluster and Open Science Grid~\cite{osg07,osg09}, supported by National Science Foundation award 1148698, and the U.S. Department of Energy's Office of Science. Data relevant to this work have been archived and can be accessed at Duke digital repository at Ref.~\citen{mfdata}.
\end{acknowledgements}

\bibliography{abbrev}

\end{document}